\documentclass[english]{nature}
\usepackage[colorlinks=true,citecolor=blue,linkcolor=magenta]{hyperref}
\usepackage{lineno}
\pdfoutput=1
\linenumbers
\usepackage{amsmath}
\usepackage{amsfonts}
\usepackage{stmaryrd}
\usepackage{amssymb}
\usepackage{graphicx}
\usepackage{dcolumn}
\usepackage{bm}
\usepackage[font={sf, small},labelfont=bf]{caption}
\usepackage{upgreek}

\usepackage{soul}
\usepackage{changes}
\begin{document}
\nolinenumbers

\title{Self-Aligned Single-Electrode Actuation of Tangential and Wineglass Modes using PMN-PT}

\author{\noindent Ozan Erturk$^{1}$, Kilian Shambaugh$^{2}$, Ha-Seong Park$^{3}$, Sang-Goo Lee$^{3}$, Sunil A. Bhave$^{1,\dagger}$}
\maketitle

\begin{affiliations}
\item OxideMEMS Lab, Purdue University, West Lafayette, IN, USA
\item Polytec Inc., Irvine, CA, USA
\item iBULe Photonics Company Ltd., Incheon 21999, South Korea

\normalsize{$^\dagger$E-mail: bhave@purdue.edu}
\end{affiliations}

\begin{abstract}
Abstract:

Considering evolution of rotation sensing and timing applications realized in Micro-electro-mechanical systems (MEMS), flexural mode resonant shapes are outperformed by bulk acoustic wave (BAW) counterparts by achieving higher frequencies with both electrostatic and piezoelectric transduction. Within the 1-30 MHz range, which hosts BAW gyroscopes and timing references, piezoelectric and electrostatic MEMS have similar transduction efficiency. Although, when designed intelligently, electrostatic transduction allows self-alignment between electrodes and the resonator for various BAW modes, misalignment is inevitable regarding piezoelectric transduction of BAW modes that require electrode patterning. In this paper transverse piezoelectric actuation of [011] oriented single crystal lead magnesium niobate-lead titanate (PMN-PT) thin film disks is shown to excite the tangential mode and family of elliptical compound resonant modes, utilizing a self-aligned and unpatterned electrode that spans the entire disk surface. The resonant mode coupling is achieved employing a unique property of [011] PMN-PT, where the in-plane piezoelectric coefficients have opposite sign. Fabricating 1-port disk transducers, RF reflection measurements are performed that demonstrate the compound mode family shapes in 1-30 MHz range. Independent verification of mode transduction is achieved using in-plane displacement measurements with Polytec’s Laser Doppler Vibrometer (LDV). While tangential mode achieves 40$^o$/sec dithering rate at 335 kHz resonant frequency, the n=2 wine-glass mode achieves 11.46 nm tip displacement at 8.42 MHz resonant frequency on a radius of 60 $\mu $m disk resonator in air. A single electrode resonator that can excite both tangential and wine-glass modes with such metrics lays the foundation for a BAW MEMS gyroscope with a built-in primary calibration stage.
\end{abstract}

\setlength{\parskip}{12pt}%
\section{Introduction}

\begin{figure} [ht]
\centering\includegraphics[width=16 cm]{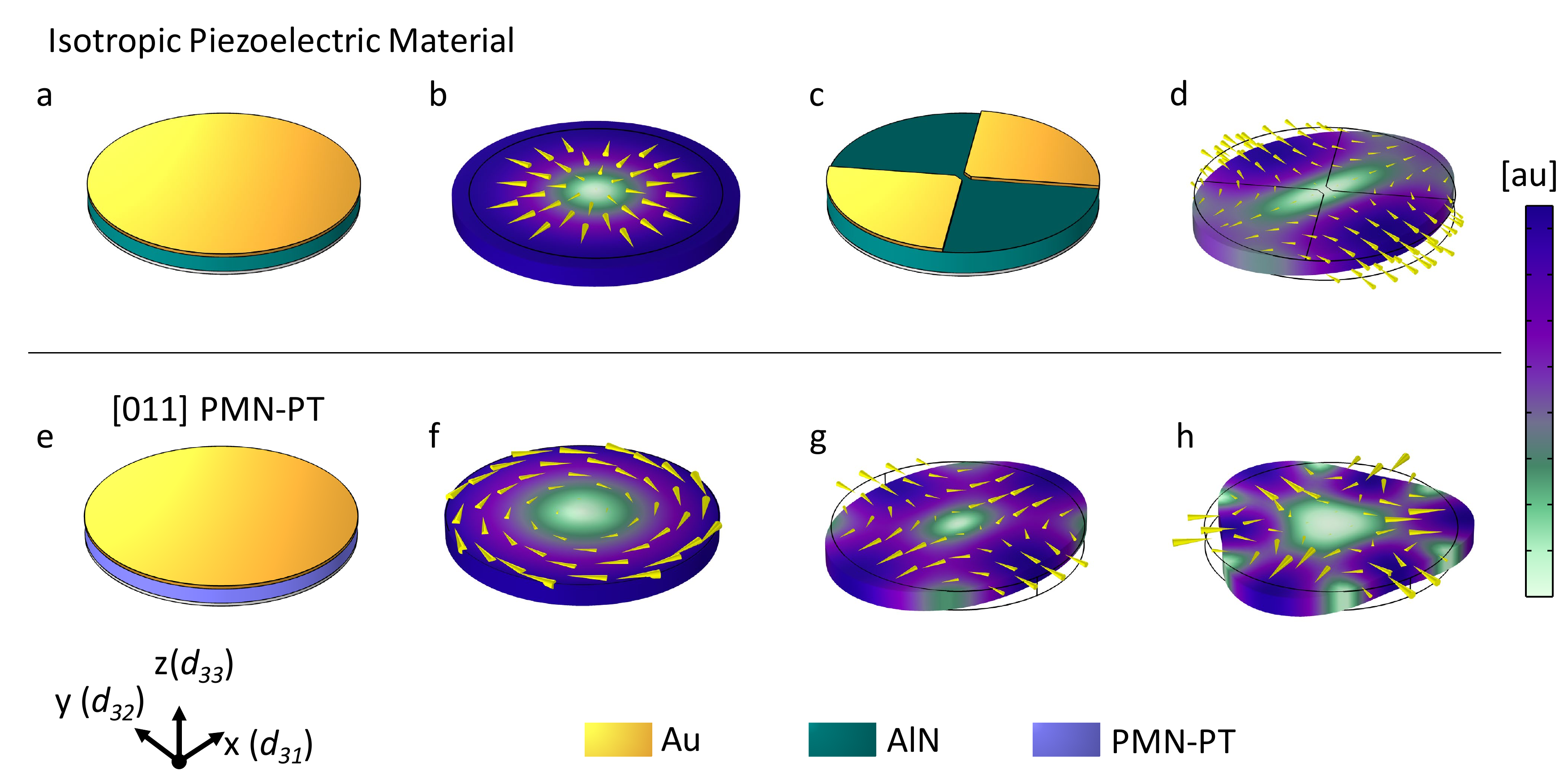}
\caption{\textbf{Schematic representation of disk resonator with single or split electrodes along with mode shapes of interest for conventional piezoMEMS materials (top row) and for PMN-PT (bottom row)}
\text Quiver plots (yellow cones that are proportional to the displacement magnitude) and color scale show displacement of each mode.
\textbf{a} Disk resonator employing conventional piezoelectric material such as AlN with unpatterned top electrode 
\textbf{b} corresponding radial mode with exaggerated deformation (due to in-plane piezoelectric coefficients ($d_{31}$ and $d_{32}$) having the same sign. 
\textbf{c} Disk resonator with segmented electrodes using the same in-plane piezoelectric coefficients with the same sign in order to excite n=2 wine-glass mode that is shown in
\textbf{d}. The exaggerated deformation of strain profiles in x- and y-direction are opposite due to segmented electrode configuration.
\textbf{e} Disk resonator employing [011] PMN-PT with a single electrode that is capable of exciting all the following mode shapes without the need for segmented electrodes due to in-plane piezoelectric anisotropy of the material:
\textbf{f} Volume conserving tangential mode shape showing shear displacement 
\textbf{g} n=2 wine-glass mode, where the maximum displacement point is along x-axis
\textbf{h} n=3 wine-glass mode. 
}
\label{Figure_1}
\end{figure}

Contour-mode bulk acoustic wave resonators have gained attention in the past decade due to their ability to beat the limitations of flexural mode resonators (such as low Q and low resonance frequency) for various application domains from RF filters to gyroscopes. Considering disk structures, displacement profiles of all contour vibrations of circular plates have been reported\cite{Onoe1956} (a correction to frequency equation to $n=2$ wine-glass mode is published in a later work\cite{Abdelmoneum2003}), where modes of vibrations are classified into two parts when circumferential order $n$ is equal to $0$: namely radial mode (absence of rotation) and tangential mode (absence of areal dilation). The higher order in-plane compound modes ($n>1$) incorporate both tangential and areal dilation at the same time. In conventional piezoelectric MEMS materials such as aluminium nitride (AlN), or lead zirconium titanate (PZT), when a disk resonator is designed with unpatterned electrodes as shown in Figure \ref{Figure_1}a, it is only possible to excite radial modes due to the strain coupling caused by the in-plane piezoelectric coefficients having the same sign. When more complex compound modes are desired for various applications, designers have to employ segmented electrodes in order to match the piezoelectrically generated strain profile to the intended contour mode of vibration and avoid charge cancellation\cite{Onoe1956,Pillai2019,Elsayed2016,Pulskamp2012a}. Figure \ref{Figure_1}b shows a schematic representation of such segmented electrode configuration that is necessary to transduce $n=2$ compound mode (wine-glass mode) as an example. This family of compound modes ($n>1$) have been demonstrated to be useful in RF filters\cite{Matsumura2010, Wei2017} and gyroscopes\cite{Hao2004,Serrano2016,Ahn2015}. However, use of segmented electrodes for mode shapes of importance such as wine-glass modes inevitably causes misalignment in piezoelectric transduction topology due to patterning of top electrodes that require a separate lithographic step during fabrication. Misalignment of the electrodes causes excitation of spurious modes that are in close proximity to the mode of interest in the frequency domain or cause undesired out-of-plane components\cite{Matsumura2010}. Self-alignment is possible in electrostatic actuation of compound modes since the same structural layer is used to realize electrodes and the resonant body\cite{Serrano2016,Pourkamali2004}. It is not possible to achieve self-alignment in the piezoelectric domain to realize any resonant mode other than the radial mode depicted in Figure \ref{Figure_1}b. Therefore, even if piezoelectric transduction is accepted to provide more robust integration and efficient coupling compared to electrostatic transduction of BAW modes, in recent works it has been reported that microfabrication advancements allowed comparable performance of BAW gyroscopes and timing references for electrostatically actuated BAW modes\cite{DeVoe2001,Dulmet2016,Serrano2016,Vukasin2020,Kaajakari2019}. As electrostatic transduction is achieved through normal forces, shear displacement required to achieve tangential mode is not possible to excite using electrostatics. Moreover, tangential mode cannot be excited with traditional piezoelectric materials because their in-plane piezoelectric coefficients have the same sign\cite{Onoe1956,Sakr2006,Pulskamp2012a}. Therefore, there has been no demonstration of a tangential mode or compound modes using a single unpatterned electrode.

In this paper, we demonstrate efficient transduction of compound modes as well as the tangential mode in a piezoelectric disk resonator by a single self-aligned electrode utilizing the in-plane anisotropy of [011] lead magnesium niobate-lead titanate (PMN-PT). We extensively report on transduction of $n=2$ wine-glass mode (WGM) because it is of particular interest for gyroscopy. In addition, we measure and optically verify the tangential mode transduction. The tangential mode is isochoric with no out-of-plane-displacement, which is an attractive feature for MEMS clocks, resonant sensors, and an integrated calibration stage for gyroscopes.

\section{Results}
\subsection{Lead magnesium niobate-lead titanate (PMN-PT)}

PMN-PT is a relaxor type ferroelectric material, discovered in 1997. It quickly became popular in various applications due to its superior piezoelectric constant and high coupling coefficient values as well as its ability to be grown in single crystal form\cite{Sun2014,Park2011}. $[011]_c$ single crystal PMN-PT is reported to have opposite polarity of the in-plane piezoelectric coefficients\cite{Liu2011}, namely $d_{31}$ and $d_{32}$ as shown in Figure \ref{Figure_1}. This unique property is utilized to rotate magnetic domains in multiferroic structures, taking advantage of the compressive stress along one direction and tensile stress along the orthogonal in-plane direction\cite{Wang2018, Chen2019}. Correspondingly, WGM shape is realized without the need for segmented electrodes (as shown in Figure \ref{Figure_1}c) since PMN-PT produces in-plane strain that has opposite direction in x- and y-axes under transverse electric field, hence enforcing wine-glass mode shape naturally. Although not as intuitive, the unpatterned electrode can excite the tangential mode on the same disk structure (as depicted in Figure\ref{Figure_1}f) as well as higher order n=3 WGM shape.

In order to investigate the transverse actuation of modes of interest and their operation of frequency, finite element analysis (FEA) modeling of the disk resonators is performed using the COMSOL piezoelectric module. Adopting the material properties of [011] PMN-PT with $d_{31}$ and $d_{32}$ opposite signs\cite{Liu2011}, a simplified COMSOL model with side tethers is implemented. This model calculates the mode shapes as shown in Figure \ref{Figure_1}f-h, and resonant frequencies of disk resonators with various radii ranging from 20 $\mu m$ to 75 $\mu m$, where resonant frequencies as a function of disk radius are plotted in Figure \ref{Figure_S11}b.

\begin{figure} [ht]
\centering\includegraphics[width=15.5 cm]{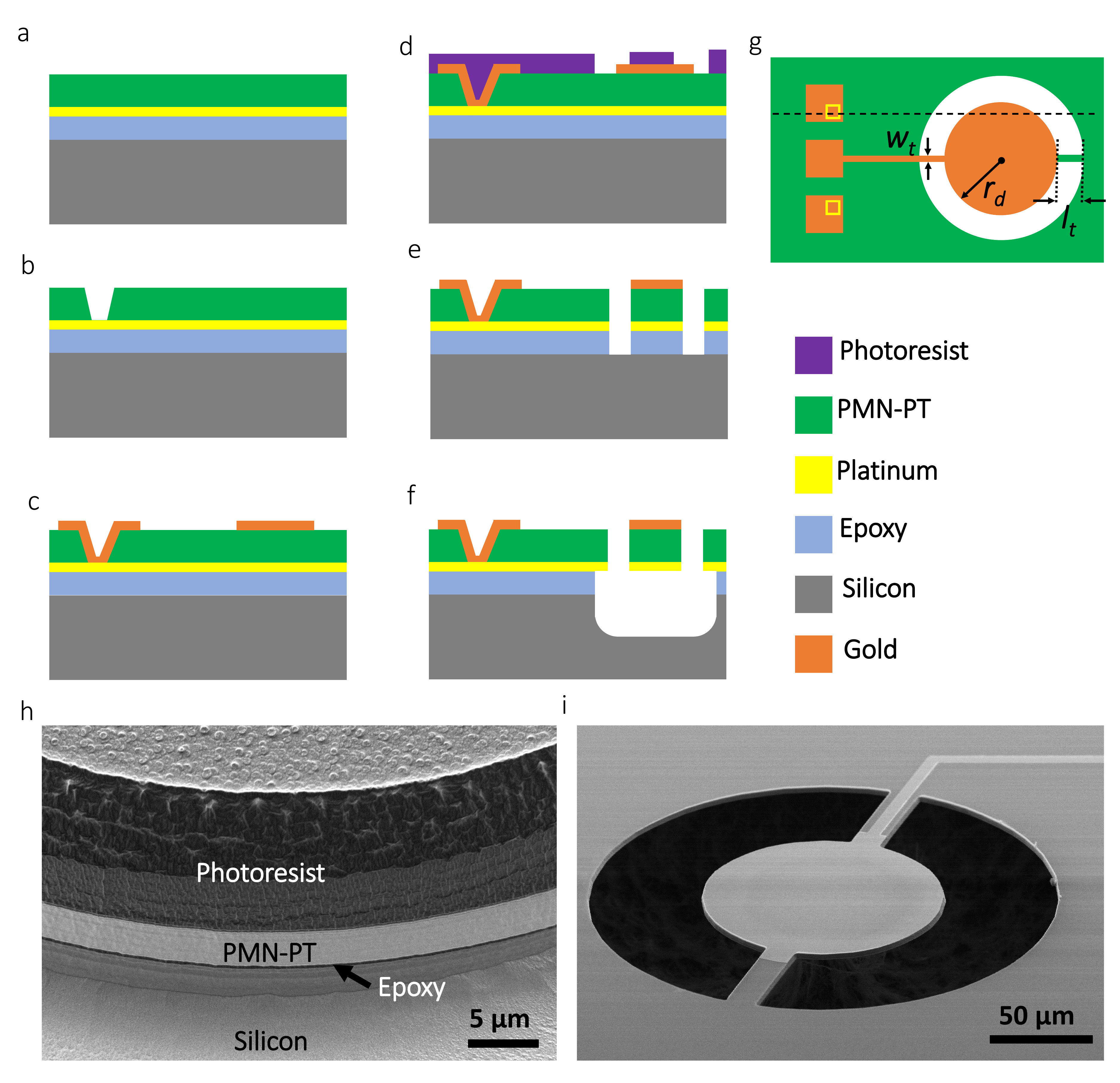}
\caption{\textbf{Fabrication process and SEM images} 
\textbf{a} Initial layer stack schematic representation.
\textbf{b} Wet etching of via openings for bottom electrode access.
\textbf{c} Top metal deposition of Au layer
\textbf{d} Lithographic patterning of etch gaps defining the resonator body smaller than the top electrode ensuring self-alignment.
\textbf{e} Ion milling (CHISEL) of resonator body along with side tethers.
\textbf{f} Releasing the resonator by XeF\textsubscript{2} etching of the silicon layer.
\textbf{g} Schematic top view of the fabricated disk resonator and geometrical definitions
\textbf{h} SEM image of the sidewall after CHISEL etch with annotated layers. Due to the cyclic changing incident ion beam, the photoresist layer seems to have different textures on the sidewall.
\textbf{i} SEM image of the suspended device with two side tethers.
}
\label{Figure_Fab}
\end{figure}

Single-crystal PMN-PT film samples with $6 \mu m$-thickness developed by iBule Photonics are bonded to silicon carrier wafers as schematically shown in Figure \ref{Figure_Fab}a. While most bonding and fabrication process steps are identical to typical single-crystal Piezo-on-Insulator (POI) resonator fabrication, the most notable differences are in Figure \ref{Figure_Fab}c and d. A larger than required metal top electrode is patterned, in Figure \ref{Figure_Fab}c. Then the disk resonator body is defined and patterned smaller than the top metal to achieve self-alignment of the top electrode to the disk using CHISEL (CHanging Incident beam-angle for Sidewall Etching and Lapping, Supplementary Note 1). An SEM image of the completed etch step is shown in Figure \ref{Figure_Fab}h with annotated layers. Finally, exposed silicon is sacrificially etched using XeF\textsubscript{2} resulting in the self-aligned electrode resonator Figure \ref{Figure_Fab}e. An SEM image of the released disk resonator is shown in Figure \ref{Figure_Fab}g.

\subsection{Device Design and Characterization}

\begin{figure} [ht]
\centering\includegraphics[width=16.5 cm]{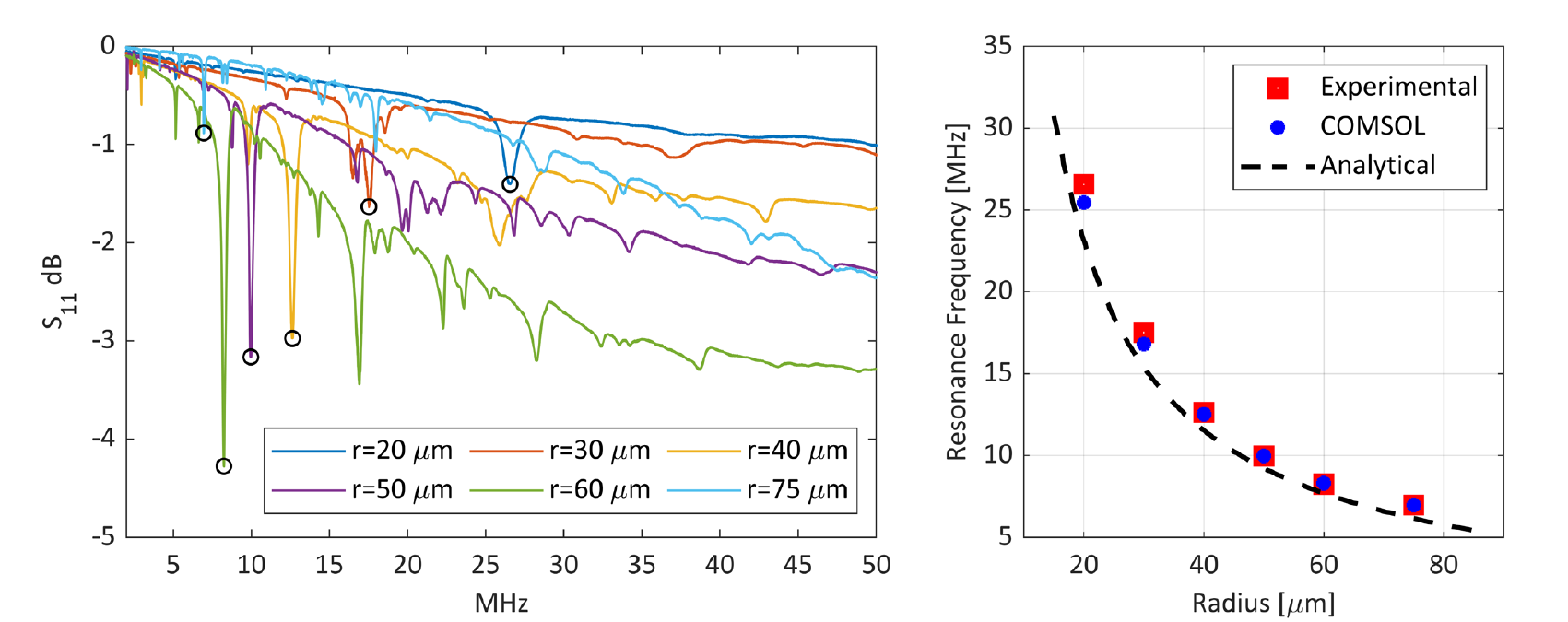}
\caption{\textbf {Electrical characterization results of disk resonators with varying radii with corresponding resonant frequencies }
\textbf {a} Overlaid plots of measured $S_{11}$ data of disks with varying radii. The resonant dip that corresponds to WGM is encircled.
\textbf {b} WGM frequency vs. disk radius, analytical, FEA, and measurement results compared. Note that the effect of anchors and tethers gets more drastic as the disk radius gets relatively smaller, hence the empirical results deviate more from the calculated values, especially for resonant frequency values estimated by analytical equations.
}
\label{Figure_S11}
\end{figure}

\begin{figure} [ht]
\centering\includegraphics[width=16 cm]{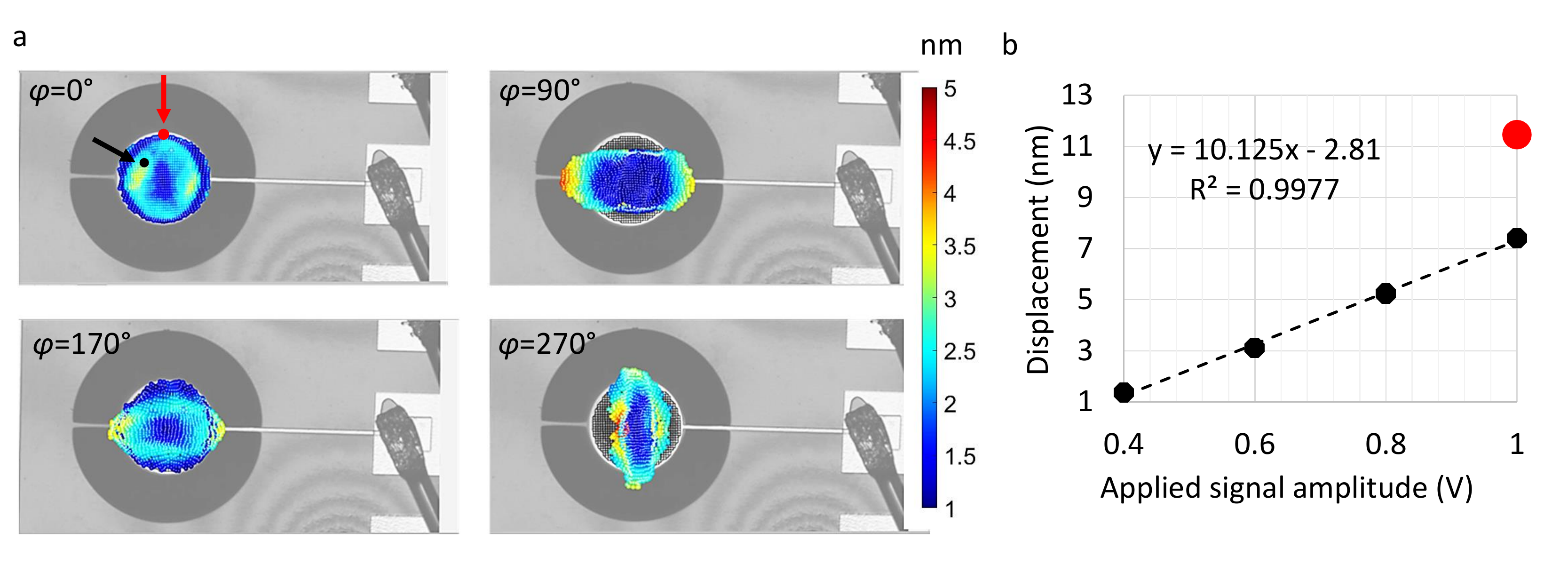}
\caption{\textbf {LDV results for n=2 wine-glass mode}
\textbf{a} LDV snapshots of a disk resonator with $r=60 \mu m$ radius shown at different phases of a single oscillation cycle at $\varphi=0^{\circ}$, $\varphi=90^{\circ}$, $\varphi=170^{\circ}$, and $\varphi=270^{\circ}$ 
\textbf{b} Displacement in the y-direction as a function of applied signal amplitude. Two distinct data points are plotted while maximum tip location is recorded (noted in red circle in the $\varphi=0^{\circ}$ phase) while an inner point (noted by a black circle in the $\varphi=0^{\circ}$ phase) displacement is recorded for various applied voltages. Linear fit to the displacement data along with the goodness of the fit is represented.
}
\label{Figure_LDV_WG}
\end{figure}

\begin{figure} [ht]
\centering\includegraphics[width=16 cm]{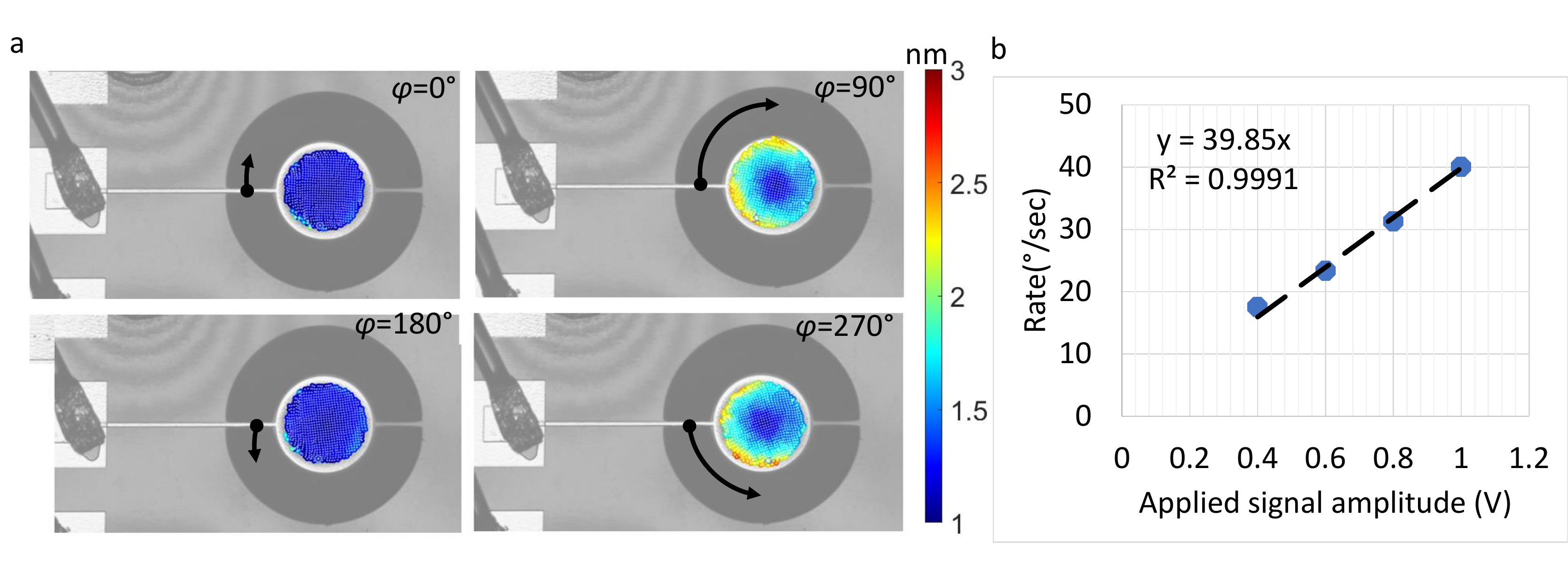}
\caption{\textbf {LDV results for tangential mode}
\text LDV snapshots of a disk resonator with $r=60 \mu m$ radius shown at different phases of a single oscillation cycle
\textbf{a} LDV snapshots of disk resonator with $r=60 \mu m$ radius shown at different phases of a single oscillation cycle at $\varphi=0^{\circ}$, $\varphi=90^{\circ}$, $\varphi=180^{\circ}$, $\varphi=270^{\circ}$ 
\textbf{b} Rate of rotation is calculated from x- and y-displacement, and plotted with respect to applied signal amplitude. Linear fit of the data points is presented.
}
\label{Figure_LDV_Tangential}
\end{figure}

Considering the most important geometric factor that determines the resonant frequency, namely radius of the disk, we designed disk resonators that have resonance frequency around 10MHz for wine-glass mode with radii ($r_d$ in Figure \ref{Figure_Fab}g) ranging from $20\mu m$ to $75\mu m$. Given the material properties and analytical equations provided\cite{Onoe1956} (a correction to n=2 wine-glass mode family frequency equations is provided\cite{Abdelmoneum2003}), estimations of radius values necessary to obtain WGM in the 1-30 MHz are determined and verified by FEA analysis. Side tethers are designed to decrease mechanical loss by choosing the lengths of the tethers ($l_t$ in Figure \ref{Figure_Fab}g) to match the the quarter wavelength of n=2 WGM of varying radii. The width of the tethers ($w_t$ in Figure \ref{Figure_Fab}g) are chosen to be as narrow as possible allowed by the critical dimension of the lithography\cite{Abdelmoneum2003,ZhiliHao2005}. 

Electrical RF reflection ($S_{11}$) measurements are performed on fabricated 1-port devices in air after the release step using Ground-Signal-Ground (GSG) pads as schematically shown in Figure \ref{Figure_Fab}g. Figure \ref{Figure_S11}a shows the RF response of disk resonators with radii values ranging from 20$\mu m$ to 75 $\mu m$. It is shown that although some spurious modes exist for some radii values, the most prominent resonant shape in the range of interest is the desired WGM. A comparison of the resonant frequencies for different radii resonators is shown in Figure \ref{Figure_S11}b. For larger radii, the agreement between calculated and experimental data is greater due to influence of the side tethers getting less significant compared to the resonant body size.

We perform an independent verification of the mode shapes using a Laser Doppler Vibrometer (LDV). LDV videos of the tangential mode and n=2,3 WGM are provided in supplementary material (Supplementary Video 1,2, and 3 respectively). Cartesian grids (as opposed to polar distribution of data points) are employed for scanning the top surface of the resonators in order to establish equal spacing between data points and eliminate any possible spatial aliasing and geometrical biasing of the circular scanned space. Snapshots of the measured WGM are shown in Figure \ref{Figure_LDV_WG}a at different phases of one cycle of oscillation for a disk resonator with radius of $60 \mu m$ at 8.42 MHz. LDV measurement and data acquisition technique is described in detail in supplementary information (Supplementary Note 2). 

We record the Y-displacement at two locations on the disk for the n=2 WGM. At Location-2, (denoted with a red-circle) which is the edge of the disk we measured a displacement of 11.46 nm in the Y-direction for an excitation amplitude of 1V as shown in Figure \ref{Figure_LDV_WG}b, which is a comparable value to that of electrostatically transduced BAW gyroscopes. Since complete mapping of each disk resonator takes about 50 minutes, the LDV measurement grid may drift with respect to the resonator over multiple scans. Therefore, in order to measure linearity, we chose Location-2 (denoted with the black-circle) that resides slightly towards the center to achieve a more stable measurement point as depicted in Figure \ref{Figure_LDV_WG}a. The measured Y-displacement at this location shows linear displacement with applied voltage amplitude for the WGM.

We also measured the tangential mode using LDV. Snapshots of a single oscillation cycle is provided in Figure \ref{Figure_LDV_Tangential}a for the same $60 \mu m$ radius device. Since LDV measures in-plane displacement in X- and Y- directions, we calculated the angular displacement along a radial line with an average displacement of $19 \mu ^{\circ}$ at 335 kHz, which is the resonant frequency of the tangential mode for a disk with $r=60 \mu m$ radius. This corresponds to a rotation rate of $40^{\circ}/s$ when a 1V amplitude excitation signal is applied. In Figure \ref{Figure_LDV_Tangential}b, the rate data is recorded for different voltage amplitude values showing a linear dependency of the rotation rate on the applied signal.

\section{Discussion}

We achieved transverse actuation of tangential and compound mode family resonant shapes on a piezoelectric disk resonator using a self-aligned single-electrode exploiting the unique in-plane anisotropy of [011] oriented single crystal PMN-PT. We developed a wafer-scale fabrication procedure involving a combination of wet etching and ion milling of PMN-PT films that is thickness non-uniformity tolerant and capable of producing a vertical side-wall profile with CHISEL technique. Considering gyroscope applications of the n=2 WGM, tip displacement of 11.46 nm is measured in LDV with 1V amplitude actuation signal. The self-aligned feature of this actuator can be attractive for quadrature-error free gyroscopes\cite{Hodjat-Shamami2020}, tangential-mode TED-free oscillators. It is also important that obtained rotation rate values of the tangential mode and its linear dependency on the amplitude of the applied signal indicates that the tangential mode itself can serve as a micro rate table for in-site scale-factor calibration of gyroscopes\cite{Aktakka2017,Pinrod2016,Pinrod2017}. Therefore, this technology provides the key building block for a BAW MEMS gyroscope with a built-in primary calibration stage.

\section{Materials and Methods}
    \subsection{Device fabrication} 
Single crystal $[011]_{c}$ PMN-PT thin film samples (20 mm x 21 mm and approximately 6 $\mu m$-thick) are provided by iBule Photonics as attached to silicon substrate by means of epoxy application between the film and the silicon substrate. The schematic cross-section of the layer stack is depicted in Figure \ref{Figure_Fab}a. Fabrication of the resonators starts with wet etching of the PMN-PT film to get electrical access to the bottom electrode Pt layer as shown in Figure \ref{Figure_Fab}b, where the Pt layer acts as a natural etch stop in a 10 \% HCl acid etch. The top electrode pattern along with the probe test pads and routing metal is deposited at the same lithographic step using liftoff. The top metal layer is  evaporated as Ti/Au (10 nm/100 nm), where a 10 nm thick Ti layer is used as adhesion layer. In this step, the top electrode of the disk is defined slightly larger to ensure that even in the presence of misalignment of the resonator body to the top electrode, the entire resonator top surface will be covered by the top metal electrode. The resonator body is defined along with the side tethers in the same lithographic step eliminating possible anchor to resonator body misalignment that would cause more spurious modes or more pronounced mechanical loss through anchors. The anisotropic etching of the resonator body is realized using ion milling until the silicon layer underneath the film is fully exposed. Different from conventional ion milling, a new ‘CHISEL’ (CHanging Incident beam-angle for Sidewall Etching and Lapping) approach was developed and described in detail (Supplementary Note 1). The etching step is divided into cycles of varying ion beam incident angle making it possible to define vertical and residue-free side walls as depicted in Figure \ref{Figure_Fab}h.

    \subsection{Electrical and Optical Measurement Setup}
$S_{11}$ measurements are performed using a ground-signal-ground (GSG) probe along with a network analyzer upon following a standard Short Open and Load (SOL) calibration protocol on an appropriate calibration substrate. Samples are wirebonded to a fan-out PCB that allows coaxial connector connectivity for excitation for LDV measurements, which are performed using Polytec MSA 100-3D. Surface modifications were implemented using photoresist patterns to enable lateral displacement detection with enhanced diffuse scattering of the laser from the top electrode surface, which otherwise provided only specular reflection due to the smoothness of the top electrode surface.

\section*{Acknowledgements}
This work was supported by DARPA MTO PRIGM-AIMS program agreement No: W911NF1820180. S.-G.L. and H.-S.P. were also supported by Korea Research Institute for Defense Technology Planning and Advancement (KRIT), F210001. The samples were fabricated in Birck Nanotechnology Center at Purdue University.

\section*{Conflict of interests}
The authors declare no competing interests.

\section*{Author contributions}
S.A.B, S.-G.L., and O.E. conceived the PMN-PT resonator design. S.-G.L. and H.-S.P. developed [011] PMN-PT-on-Silicon wafer-scale technology. O.E. performed design of experiments, microfabrication, electrical characterization and sample preparation for Laser Doppler Vibrometer (LDV) measurements. K.S. performed LDV measurements. S.A.B and S.L. supervised the project. O.E. and S.A.B. contributed to writing the manuscript with assistance from all other authors.

\section*{Additional information}
See Supplementary Information for supporting content. 

\section*{Data availability}
Data Availability Statement: The code and data used to produce the plots within this work will be released on the repository Zenodo upon publication. Correspondence and requests for materials should be addressed to S.A.B.
\clearpage
\spacing{1}

\section*{References}
\bigskip
\bibliographystyle{naturemag}
\bibliography{Ref} 

\end{document}